\documentclass[prl,aps,twocolumn,english]{revtex4}
\usepackage{graphicx}
\usepackage{subfigure}
\usepackage{amssymb}

\begin{document}

\title{Elastic theory of
Normal-Superfluid Boundary in trapped 
Fermi Gases}

\author{Stefan S. Natu}
\email{ssn8@cornell.edu}
\author{Erich J. Mueller}
\email{em256@cornell.edu}
\affiliation{Laboratory of Atomic and Solid State Physics, Cornell University, Ithaca, New York 14853, USA.}

\begin{abstract}
By modeling the normal-superfluid boundary in a trapped polarized Fermi
gas as an elastic membrane, we calculate the atomic density profile. For
small trapping anisotropy, we find that the superfluid-normal boundary remains approximately elliptical, and has an aspect ratio different from that of the trap. For very prolate clouds the boundary
becomes distorted into a capsule-like shape. 
We present an analytic explanation of this shape.
\end{abstract}
\maketitle

Despite the small numbers of particles ($N\sim 10^5$), theoretical explanations of cold atom experiments almost universally  involve only bulk properties \cite{baympethick,pethickbook}.  One exception has been experiments on two-component Fermi gases in elongated ``cigar shaped" traps \cite{hulet1,hulet2}, where in addition to bulk physics, one must also include surface effects \cite{caldas,imambekov,randeria}. Here we present a phenomenological model of these surface effects, producing a simple explanation of the observed density profiles.

In the experiments of interest, $^7$Li atoms are pumped into two internal states ($\uparrow$ and $\downarrow$) and trapped in an anisotropic harmonic trap.  Spin relaxation can be ignored in these experiments, so the number of atoms in each state $N_{\uparrow/\downarrow}$ is conserved.  The attractive short-range interactions are tuned near {\em unitarity}, where the scattering length is infinite, yielding a scale-free interaction and universal thermodynamics \cite{hohydro}.  At low temperature these gases form a superfluid where $\uparrow$ atoms pair up with $\downarrow$ atoms.  
 When the ratio of spins $N_\uparrow/N_\downarrow$ deviates from unity, the system phase separates into a central paired region surrounded by a predominately polarized region.  We present a theory of this phase separated gas.
 
Our analysis goes beyond  the standard  {\em local density approximation} (LDA) which, as described here,  can be derived from a hydrodynamic theory. Including only bulk physics,  local hydrodynamic equilibrium requires that the pressure in the trap obeys $\nabla P=- n \nabla V$ where $n$ is the local density and $V$ is the trapping potential.  Assuming an isothermal trap, this then requires that $\nabla \mu=-\nabla V$, where $\mu=(\mu_\uparrow+\mu_\downarrow)/2$ is the average chemical potential.  In the absence of spin-dependent forces, the chemical potential difference $h=(\mu_\uparrow-\mu_\downarrow)/2$ is independent of space. A consequence of this hydrodynamic assumption is that all properties of the gas are unchanged as one moves along contours of fixed $V$.  Experiments at Rice see a violation of this requirement \cite{hulet1}: the domain wall separating the superfluid and normal region does not follow an elliptical isopotential contour, rather it is ``squished" axially, forming more of a capsule shape.
 
 As discussed in \cite{theja,stoof}, this distortion is consistent with a phenomenological theory where one includes surface tension in the domain wall.  In \cite{theja}, this phenomenological theory was explored using a variational ansatz where the domain wall was taken to form an ellipse whose aspect ratio was a variational parameter.  Haque and Stoof \cite{stoof} further explored the shape of the domain wall by parameterizing it in cylindrical coordinates ($\rho,z,\phi$) as a curve obeying $(\rho/R)^\gamma+(z/Z)^\gamma=1$ where $\gamma,R,$ and $Z$ are determined variationally.   Here we optimize the shape of the domain wall without restricting its shape in any way.  We produce an analytic argument which explains the shape of the domain wall.  The accuracy of our approximations are verified via more sophisticated numerical calculations.
 
Treating the superfluid-normal interface as an elastic membrane, 
hydrodynamic equilibrium requires \cite{isenberg}
 \begin{equation}
2\sigma_{0}H=\Delta P,\label{eq:1}\end{equation}
where $\sigma_{0}$ is the surface tension, $H=(1/2)(1/R_1+1/R_2)$
is the mean curvature, expressible in terms of the principal radii of curvature $R_{1/2}$, and $\Delta P$ is the pressure difference between the superfluid and normal gas. 

In our calculations, we treat
$\sigma_{0}$ as a spatially uniform parameter, which only depends on the
average chemical potentials of the superfluid and normal gas at the
center of the trap. In a more sophisticated theory, $\sigma_{0}$ should depend on the local density at the boundary, the pressure drop $\Delta P$, and the temperature.  Given that we use an approximate equation of state, we feel that our model of the bulk properties is not sufficiently accurate to warrant including these dependencies whose quantitative forms are not all known.

As introduced by Chevy \cite{chevy}, for the bulk equation of state in the superfluid (S) and normal state (N), we use 
\begin{equation}
P_{S/N}=\left(\frac{-2}{15\pi^{2}}\right)\left(\frac{2m}{\hbar^{2}}\right)^{\frac{3}{2}}\xi_{S/N}\mu_{S/N}^{\frac{5}{2}},\label{eq:2}\end{equation}
where $\xi_{S}=\frac{1}{(2\xi)^{3/2}},$ $\xi_{N}=1$ and $\xi\approx0.45$
is a universal parameter. The chemical potentials are $\mu_{S}=\mu$
and $\mu_{N}=\mu_\uparrow=\mu+h$. This equation of state is exact for the zero temperature superfluid where the polarization is zero ($n_\uparrow=n_\downarrow$) -- it is however only approximate in the normal state where it assumes that the local polarization is $p=(n_\uparrow-n_\downarrow)/(n_\uparrow+n_\downarrow)=100\%$.

Although experiments at MIT \cite{ketterlepd} are consistent with the local polarization of the zero temperature normal state being as low as $p=35\%$, the Rice experiments \cite{hulet1} that we are mainly concerned with, find that  $p\geq95\%$.  We are unaware of an explanation for this difference in behavior.

In the superfluid at unitarity the only microscopic energy-scale is the chemical potential, while the trap provides a macroscopic length-scale $R_{TF}=\sqrt{2\mu_0/m\omega_\rho^2}$ defined in terms of the central chemical potential $\mu_0$, and the trap $V\left(\rho,z\right)=(1/2)(m\omega_{\rho}^{2}\rho^{2}+m\omega_{z}^{2}z^{2})$.  We therefore introduce a dimensionless surface tension 
\begin{equation}
\sigma=\frac{\sigma_{0}}{\left(\frac{1}{15\pi^{2}}\right)\left(\frac{2m}{\hbar^{2}}\right)^{\frac{3}{2}}\mu_{0}^{\frac{5}{2}}R_{TF}}.\label{eq:3}
\end{equation}
The anisotropy of the trap will be parameterized in terms of the dimensionless parameter $\Lambda=\omega_z/\omega_\rho$.  A prolate cigar-shaped trap has $\Lambda<1$.

We parametrize the surface by the function $z(\rho)$, whence the
 mean curvature
has the form $H=z_{\rho\rho}/\left(1+z_{\rho}^{2}\right)^{\frac{3}{2}}$,
where the subscripts refer to derivatives with respect to the radial
coordinate. 

\begin{figure}
\begin{minipage}{1\columnwidth}
\begin{tabular}{cc}
\begin{minipage}[b]{0.60\columnwidth}
\parbox[b]{0.6\columnwidth}{
(a)\includegraphics[width=0.65\columnwidth] {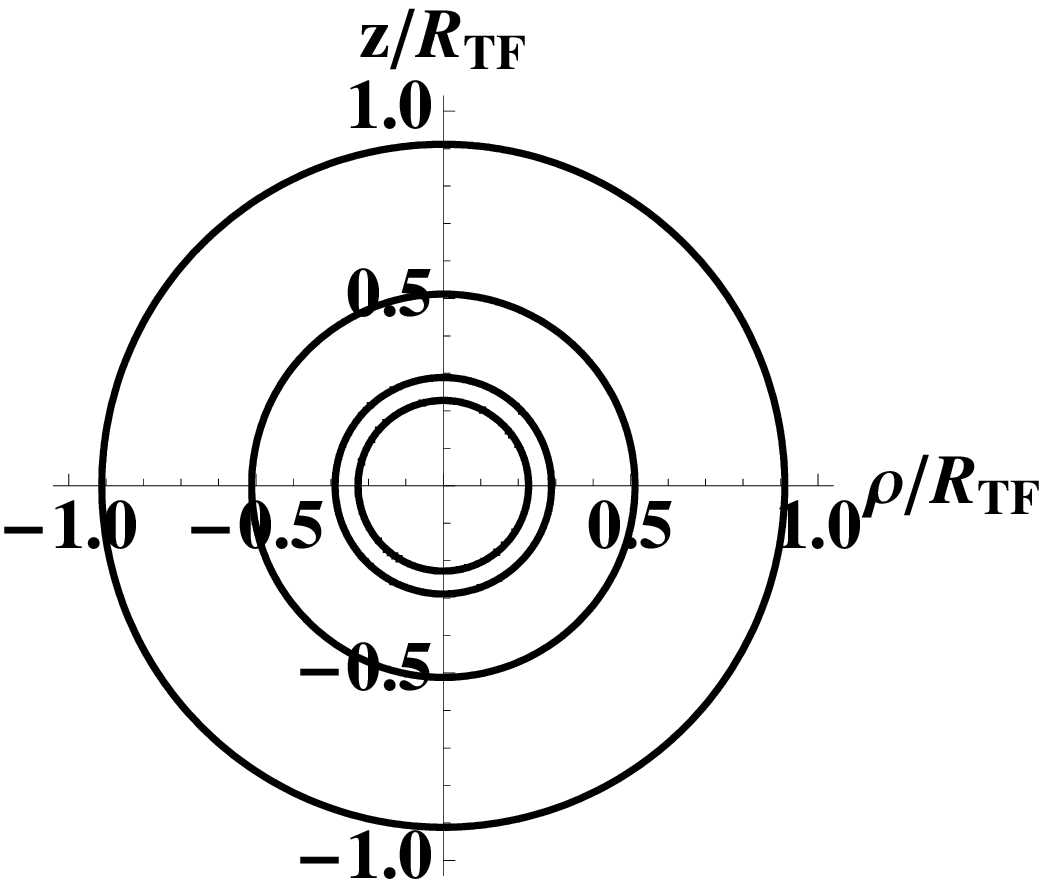}}\\
\parbox[b]{0.6\columnwidth}{
(b)\includegraphics[width=0.65\columnwidth]{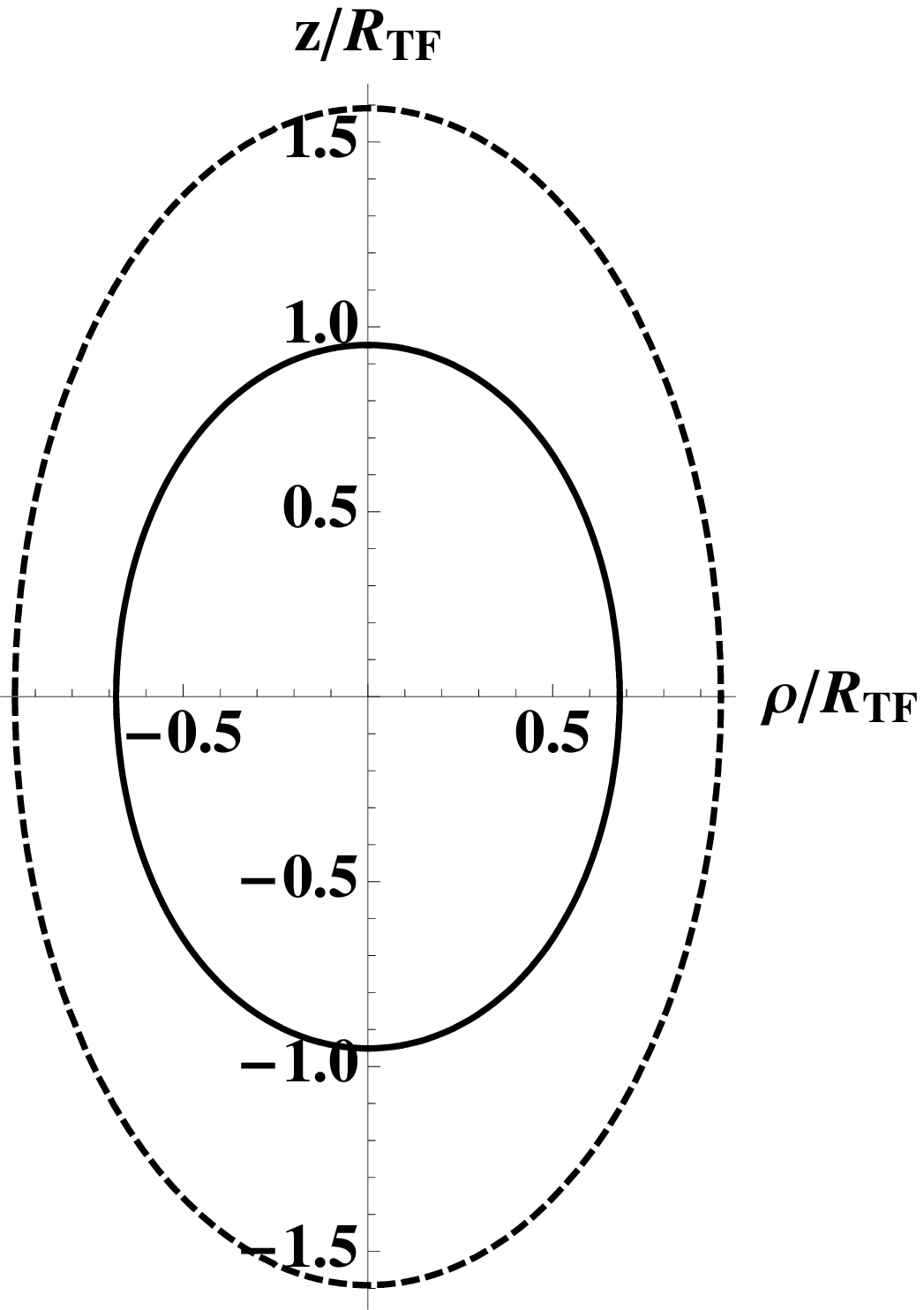}}\\
\parbox[b]{0.6\columnwidth}{
(c)\includegraphics[width=0.65\columnwidth]{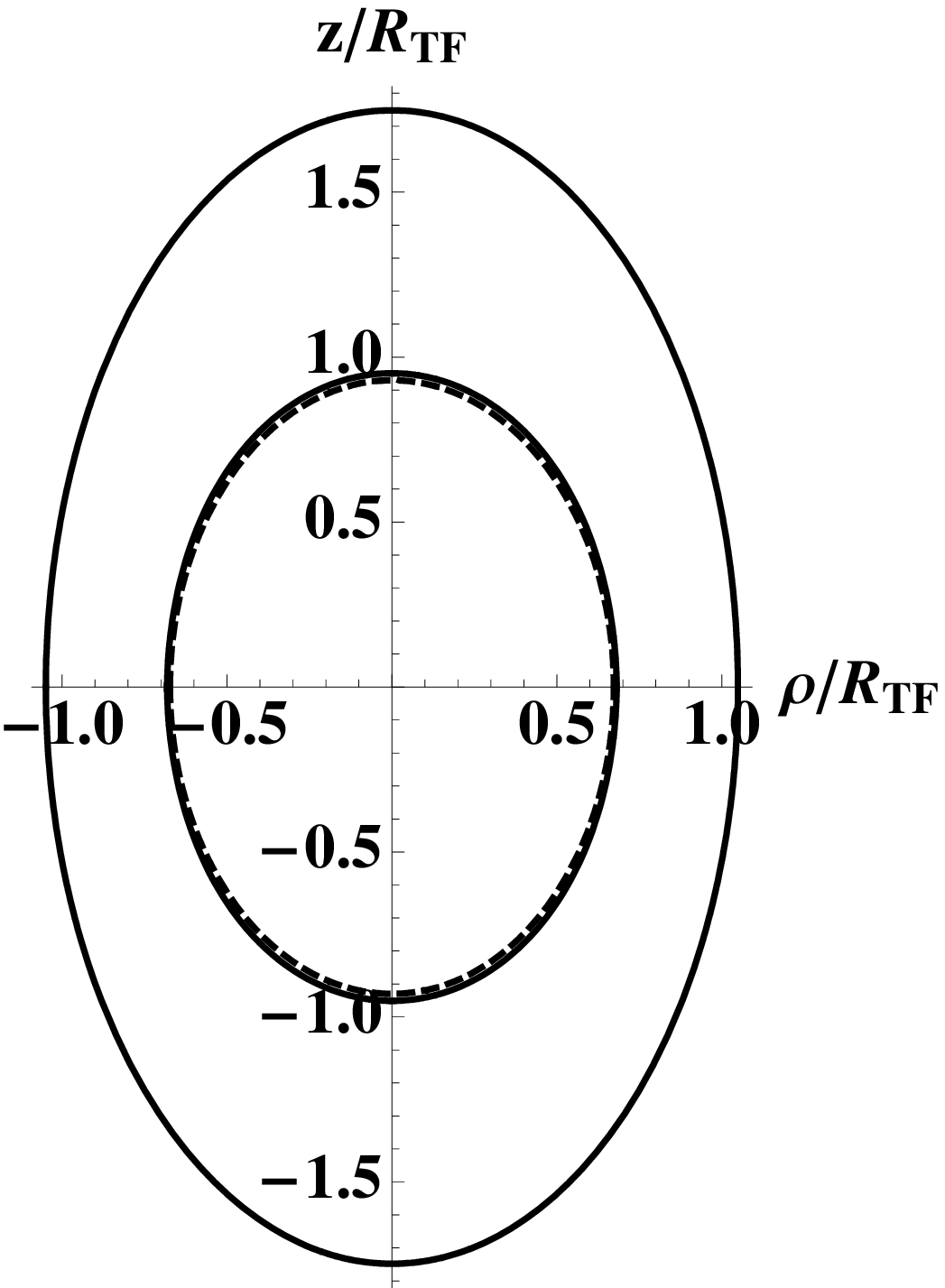}}
\end{minipage}&
\parbox[b]{0.5\columnwidth}{
(d)\includegraphics[width=0.43\columnwidth]{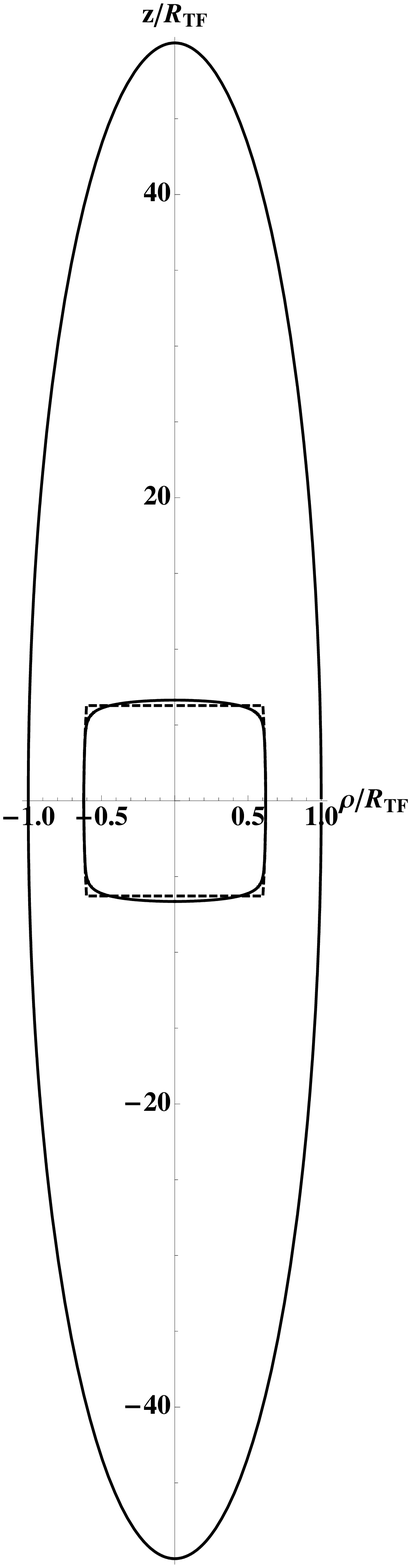}}
\end{tabular}
\end{minipage}

\caption{\label{fig:-1}  Depiction of domain walls between superfluid and normal region. For all figures, $\bar h$ equals $0.1$. 
Top Left (a): Isotropic trap. Moving inwards from
outside, the dimensionless surface tension $\sigma$ takes on the values 0, 0.3, 0.6
and 1. The radii are normalized with a non-interacting Thomas-Fermi
radius ($R_{TF}=\sqrt{2\mu_0/m\omega_\rho^2}$).  Middle left (b): Comparison of the domain wall for $\sigma=0$ (dashed curve) and $\sigma=0.5$ (solid curve) for $\Lambda= \omega_{z}/\omega_{\rho}=0.6$. Bottom Left (c): Elongated trap with $\Lambda=0.6$ and $\sigma=0.5$. The inner dashed line is the location of the boundary with the elliptical ansatz. Right (d): Cigar shaped trap with $\Lambda=0.02$ and $\sigma = 0.95$.  Inner dashed line: location of boundary from the analytic model. For both (c) and (d), the outer solid line is the extent of the atomic cloud, and the inner solid line is the shape of the boundary obtained from numerical calculations. } 
\end{figure}

With this parametrization, Eq.~\ref{eq:1} becomes
\begin{eqnarray} \label{eq:4}
2\sigma H= 2\sigma\frac{z_{\rho\rho}}{\left(1+z_{\rho}^{2}\right)^{^{3/2}}}=\frac{2}{\zeta^{\frac{3}{2}}}\left[1-\left(\rho^{2}+\Lambda^{2}z^{2}\right)\right]^{5/2}\\\nonumber-\left[1+\bar h-\left(\rho^{2}+\Lambda^{2}z^{2}\right)\right]^{5/2}\end{eqnarray}
where $\bar h=h/\mu_0$.  Solving this differential equation, with the constraint that the solution is a closed curve, determines the shape of the domain wall separating the normal and superfluid regions.

We first consider the case of an isotropic trap ($\Lambda=1$), where
contours of constant pressure drop are circles of radius $R$. In
this ansatz, the differential equation reduces to an algebraic equation
\begin{equation}
2\sigma\frac{1}{R}=\frac{2}{\zeta^{\frac{3}{2}}}\left(1-R^{2}\right)^{5/2}-\left(1+\bar h-R^{2}\right)^{5/2},\label{eq:5}\end{equation}
which is readily solved graphically.  For sufficiently large $\sigma$ (at fixed $\bar h$) there are no solutions to this equation, resulting in  a homogeneous cloud: either all superfluid or all normal.  For smaller $\sigma$ there are two solutions: one, at smaller radius, representing a local maximum of the free energy, and another at larger radius representing the local minimum.  The local maximum corresponds to the critical droplet associated with nucleation of the superfluid phase.  The local minimum is shown in Fig.~\ref{fig:-1}(a).  As one would expect, at fixed $\mu_{0}$ and $\bar h$, increasing surface tension shrinks the radius of the superfluid region.

Next, we consider the case of small trapping anisotropy, where we find that the full numerical solution to Eq.~\ref{eq:4} yields a nearly elliptical domain wall.  Assuming an elliptical ansatz, we can calculate the semi-major and semi-minor axes $a$ and $b$.  
The curvature of an ellipse
at the axial maximum is  $H_z=b/{a^{2}},$ while the curvature at the radial maximum is $H_\rho=(1/2)(a/b^{2}+1/a)$.
Substituting these expressions into Eq.~\ref{eq:4}, evaluated at
the axial and radial maxima respectively, yields two algebraic equations
that are solved for $a$ and $b$.
The results are plotted for $\bar h = 0.1$, $\sigma=0.5$ and $\Lambda=0.6$.  in Fig.~\ref{fig:-1}(b and c). The ellipse ansatz indeed works very well even for large surface tensions, and moderately anisotropic traps. However, further numerical investigation indicates that the numerical solution deviates noticeably from the ellipse ansatz when $\Lambda < 0.5$.
\begin{figure}
\begin{minipage}{1\linewidth}
\centering
\parbox[b]{1\columnwidth}{
(a)\includegraphics[width=1\columnwidth]{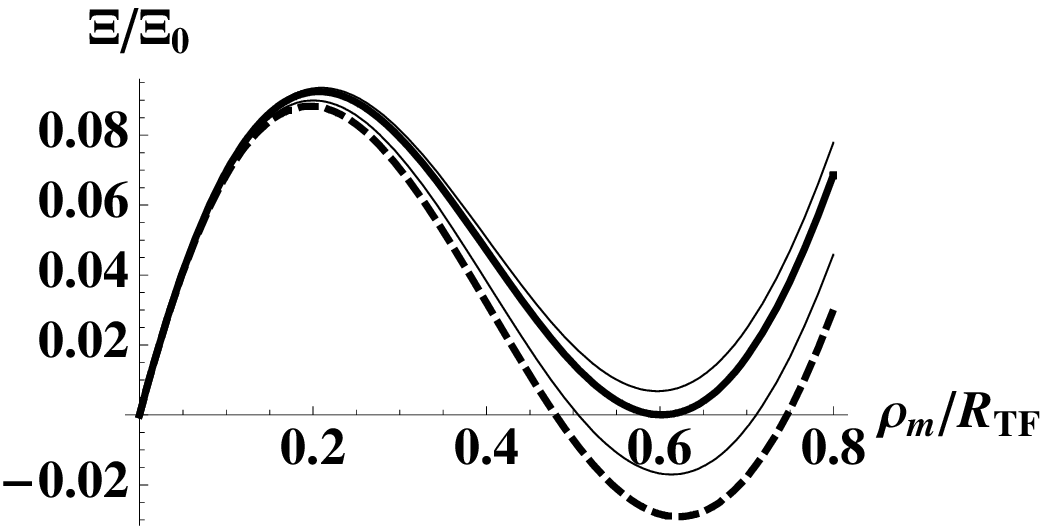}}
\hspace{0.01\textwidth}
\parbox[b]{1\columnwidth}{
(b)\includegraphics[width=1\columnwidth]{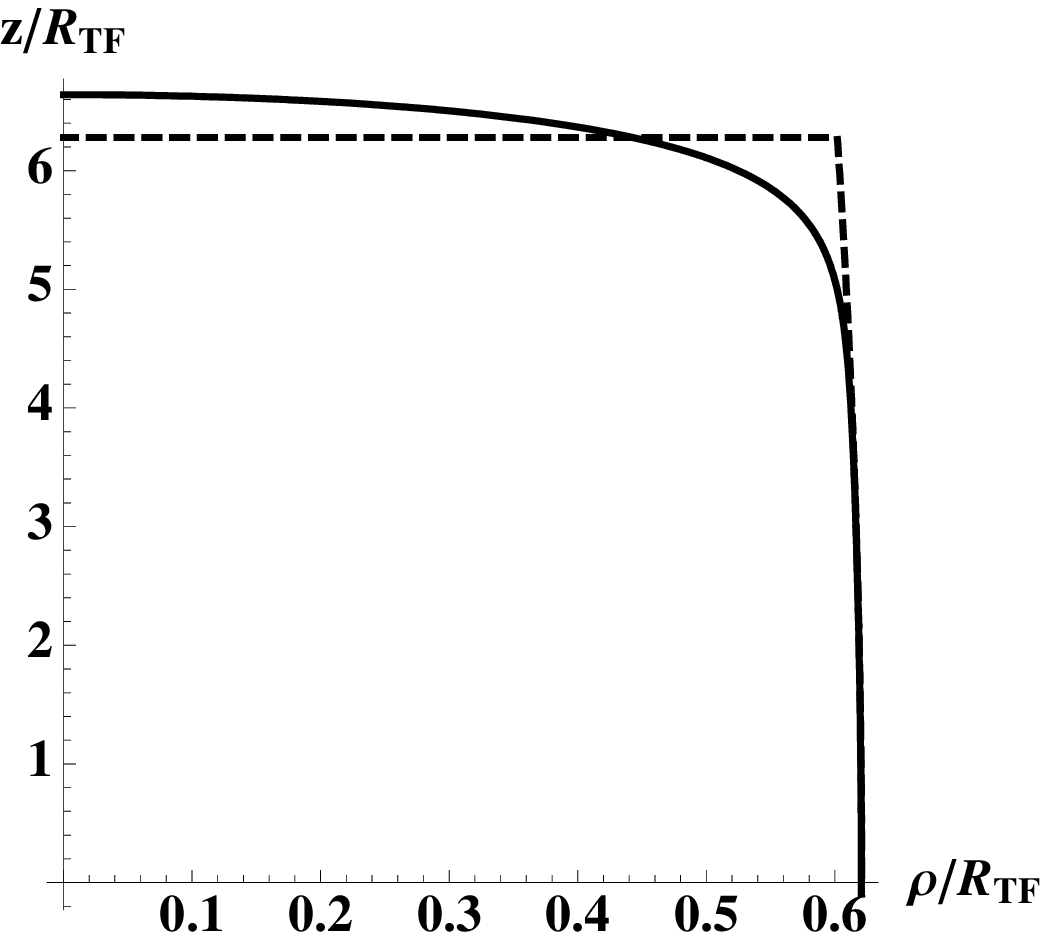}}
\end{minipage}
\caption{\label{fig:2}Top (a):  Free energy per unit length, {\small $\Xi\left(z\right)-\Xi\left(0\right)$, shown as a function of the radius of the domain wall for various
values of the axial position $z$ for $\Lambda=0.02$, $\sigma=0.95$ are shown. The
dashed curve is the free energy for $\Lambda=0$ and $\sigma=0.95$.
The thick curve plotted for $z/R_{TF}=6.28$, is where the free energy
minimum occurs at $\rho = \rho_{m}(z_{max}) = 0.62R_{TF}$ and $\rho=0$. The free energy per unit length $\Xi$ is
normalized with respect to $\Xi_{0}=\left(\frac{1}{15\pi^{2}}\right)\left(\frac{2m}{\hbar^{2}}\right)^{\frac{3}{2}}\mu_{0}^{\frac{5}{2}}R_{TF}^{2}$
.Bottom (b): A quadrant of the numerically determined
shape for $\Lambda=0.02$ and $\sigma=0.95$. The dashed line is the
shape predicted from the analytic model for the same parameters.
The numerical maximum occurs at $z=6.56 R_{TF}$ and the theoretical
maximum occurs at $z=6.28 R_{TF}$. The value of $\bar h$ was set to $0.1$ in all these calculations. 
} } 
\end{figure}

For large anisotropies $\Lambda\ll1$, the boundary becomes distinctly non-elliptical, with a pronounced ``flattenning" on the axial ends.  In the extreme case, illustrated in Fig.~\ref{fig:-1}(d), the domain wall becomes ``boxy" with apparently sharp corners.  If one parameterized the boundary by writing the radial coordinate as a function $\rho_m(z)$, this function appears to be nearly discontinuous. This boxy shape has been seen in experiments performed at Rice
\cite{hulet1}, and in  variational calculations \cite{stoof}. 

We explain this shape by writing the free energy of the system as 
\begin{eqnarray}\label{eq:6}
\Phi&=&2\pi\int dz \Xi(z)\\\nonumber
\Xi(z)&=&
-\int_{0}^{\rho_{m}}\!\!\rho d\rho P_{S}
-\int_{\rho_{m}}^{\rho_{e}}\!\!\rho d\rho P_{N} +\sigma\rho_{m}\sqrt{1+\left(\frac{d\rho_{m}}{dz}\right)^{2}},
\end{eqnarray} 
where the dependence of $\rho_{m}$ and $\rho_{e}$ on $z$ is implicit. The bulk free energy of the system has contributions from the superfluid and the normal gas. For every value of $z$, $P_{S}$ is integrated outward from $0$ to the domain wall
($\rho_{m}(z)$), and $P_{N}$  is integrated from the domain wall to the edge of the trap ($\rho_{e}(z)$). 
The contribution from the surface free energy is  $\sigma dA$,
where the differential area $dA$ is expressed as
$\rho_{m}\sqrt{1+\left(\frac{d\rho_{m}}{dz}\right)^{2}}dz$.
The angular term has been integrated out. 

In the large aspect ratio limit, the slope $d\rho/dz$ of the boundary should almost everywhere be small.  To lowest order we neglect this term, and $\Xi$ becomes only a function of $\rho_m(z)$ and not its derivative.  Thus to minimize $\Phi$, we simply need to separately minimize $\Xi(z)$ for each $z$. In Fig.~\ref{fig:2}(a) we plot $\Xi$ as a function of $\rho_m(z)$ for various values of $z$.  The near discontinuity observed in our numerical calculation is understood by noting that for a particular $z$, $\Xi$ can have multiple minima. Thus when the derivative terms are neglected, the $\rho_m(z)$ will discontinuously change from a finite value ($\rho_{m}(z_{max})$) to zero when one passes the location where these two minima have the same energy (for example, see the thick curve in Fig.~\ref{fig:2}(a)).  This discontinuity is analogous to the physics of a first order phase transition.

\begin{figure}
\begin{minipage}{1\linewidth}
\centering
\parbox[b]{1\columnwidth}{
\includegraphics[width=\columnwidth]{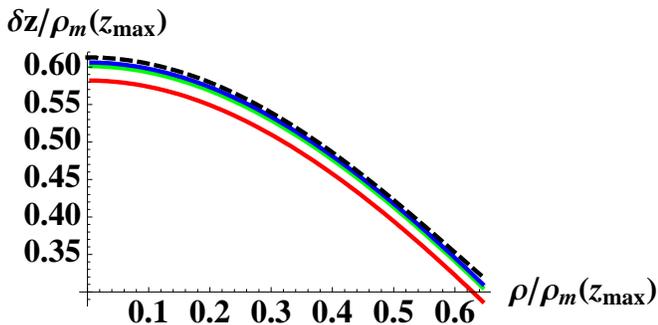}}
\end{minipage}
\caption{\label{fig:3} (Color Online) Difference between the numerical shape of the domain wall and the lowest order analytic model ($\delta z (\rho)$) away from the sharp corners, for various values of $\Lambda$ at fixed $\bar h (=0.1)$ and $\sigma (=0.95)$. From bottom to top, $\Lambda = 0.02, 0.01$, and $0.0025$. As $\Lambda \rightarrow 0$, the difference between the numerical and analytically obtained axial maximum ($\delta z^{(0)} $)$= 0.61\rho_{m}(z_{max})$ ($= 0.38R_{TF}$). Note that $\frac{\delta z^{(0)}}{\rho_{m}(z_{max})} \approx {\cal O}(1)$ in agreement with the variational argument.The dashed line is the first order perturbative calculation of $\delta z(\rho)$ for $\bar h =0.1$, $\Lambda=0$ and $\sigma=0.95$, which almost entirely accounts for the difference between the analytic and numerical results. }
\end{figure}
 
We compare the shape predicted by this analytic model to the one found in our numerics.
As $\Lambda \rightarrow 0$, the error ($\delta z^{(0)}$)  between the numerically and theoretically obtained axial maxima ($z_{max}$)  approaches a value that depends on $\sigma$ and $\bar h$, and scales as $\rho_{m}(z_{max})$. This scaling can be understood by a variational argument. 

Near the axial maximum, one can expand the radial coordinate $\rho_{m}(z)$ as
 \begin{equation}\label{eq:7}
\rho_{m}(\delta z) = \rho_{m}(z_{max}) - \frac{\rho_{m}(z_{max})}{(\delta z^{(0)})^{2}}(\delta z)^{2}.
\end{equation}
In terms of the new variables $\delta\bar z = \frac{\delta z}{\delta z^{(0)}}$ and $\bar\rho = \frac{\rho}{\rho_{m}(z_{max})}$, the free energy (Eq.~\ref{eq:6}) takes the form 
\begin{equation} \label{eq:8}
\Phi(\delta z^{(0)})=2\pi\delta z^{(0)}\int_{0}^{1}d\delta\bar z~\left(\Xi_{b}(\delta\bar z)+\Xi_{s}(\delta\bar z)\right),
\end{equation}
where the free energy density $\Xi(\delta\bar z)$ has contributions from the bulk fluids ($\Xi_{b}$) and the domain wall ($\Xi_{s}$).
The bulk energy density is
\begin{equation}\label{eq:9}  
\Xi_{b} =(\rho_{m}(z_{max}))^{2}\left(\int_{1-(\delta \bar z)^{2}}^{1}\bar \rho d\bar \rho P_{S}-\int_{1-(\delta \bar z)^{2}}^{1}\bar\rho d\bar\rho P_{N}\right),
\end{equation}
and the surface energy density is
\begin{equation}\label{eq:10}
\Xi_{s} = \sigma\rho_{m}(z_{max})\left(1-(\delta \bar z)^{2}\right)\sqrt{1+\left(\frac{\rho_{m}(z_{max})}{\delta z^{(0)}}\right)^{2}(\delta\bar z)^{2}},
\end{equation}
where we use the fact that the surface energy density has the form $\rho_{m}\sqrt{1+\left(\frac{d\rho_{m}}{dz}\right)^{2}}$. Upon integrating the free energy densities in Eq.~\ref{eq:8}, we find the free energy has a minimum, whose location depends on the specific choice of $\sigma$ and $\bar h$, and scales as $\frac{\delta z^{(0)}}{p_{m}(z_{max})}  \approx {\cal O}1$. We have confirmed this scaling by comparing the numerical and analytically obtained axial maxima for a range of $\sigma$ and $\bar h$ values.
 
Furthermore, as $\Lambda \rightarrow 0$, the relative error  ($\delta z^{(0)}/z_{max}$) scales as $\Lambda$. We can investigate the shape of the domain wall in the region near $\rho=0$ by a perturbative expansion. For $\Lambda=0$, we write $\delta z(\rho) = z(\rho) - z_{max} = \delta z^{(0)} + \delta z^{(1)} \rho^{2}+\delta z^{(2)} \rho^{4}$, where we have used cylindrical symmetry to set the coefficients of the odd powers to zero. Expanding both sides of Eq.~\ref{eq:4} in powers of $\rho$, one can solve for the coefficients $\delta z^{(1)}$, and $\delta z^{(2)}$. We plot $\delta z(\rho)$ for various values of $\Lambda$  in Fig.~\ref{fig:3} and show that after including the first order corrections, the analytic calculation is quite close to what we found numerically. 

To summarize, we have determined the shape of the superfluid-normal gas domain wall for a range of trapping anisotropies and surface tensions. For small trapping anisotropies, the shape is described by a simple elliptical ansatz. For large trapping anisotropies, like those in the experiments at Rice \cite{hulet1}, the domain wall is boxy, with sharp corners. We have developed a simple model to explain this shape, and verified our results with numerical calculations. 

We would like to thank Theja N. De Silva for useful discussions, and for providing comparisons with his numerical data. We also acknowledge relevant discussions of experimental issues with Randall Hulet, Wenhui Li and Wolfgang Ketterle. This work was supported in part by the National Science Foundation through grant PHY-0456261.

\end{document}